\documentclass[12pt]{article}
\usepackage{amsmath,epsf,amssymb,cite}
\newtheorem{theorem}{Theorem}

\newif\iffigs\figstrue

\DeclareFontFamily{U}{rsf}{}
\DeclareFontShape{U}{rsf}{m}{n}{
  <5> <6> rsfs5 <7> <8> <9> rsfs7 <10-> rsfs10}{}
\DeclareMathAlphabet\Scr{U}{rsf}{m}{n}

\def\pplogo{\vbox{\kern-\headheight\kern -29pt
\halign{##&##\hfil\cr&{
\ppnumber}\cr\rule{0pt}{2.5ex}&\ppdate\cr}
}}
\makeatletter
\def\ps@firstpage{\ps@empty \def\@oddhead{\hss\pplogo}%
  \let\@evenhead\@oddhead 
}
\def\maketitle{\par
 \begingroup
 \def\thefootnote{\fnsymbol{footnote}}
 \def\@makefnmark{\hbox{$^{\@thefnmark}$\hss}}
 \if@twocolumn
 \twocolumn[\@maketitle]
 \else \newpage
 \global\@topnum\z@ \@maketitle \fi\thispagestyle{firstpage}\@thanks
 \endgroup
 \setcounter{footnote}{0}
 \let\maketitle\relax
 \let\@maketitle\relax
 \gdef\@thanks{}\gdef\@author{}\gdef\@title{}\let\thanks\relax}
\makeatother

\def\C{{\mathbb C}}
\def\P{{\mathbb P}}
\def\Q{{\mathbb Q}}

\def\Z{{\mathbb Z}}

\def\Hom{\operatorname{Hom}}
\def\Ext{\operatorname{Ext}}
\def\Tors{\operatorname{Tors}}
\def\Free{\operatorname{Free}}

\def\SU{\operatorname{SU}}
\def\GU{\operatorname{U{}}}

\def\Spin{\operatorname{Spin}}

\def\su{\operatorname{\mathfrak{su}}}

\def\CY{Calabi--Yau}

\def\MW{Mordell--Weil}

\def\cM{{\Scr M}}

\def\cD{{\Scr D}}

\def\cX{{\Scr X}}
\def\cG{{\Scr G}}

\def\cMc{{\hfuzz=100cm\hbox to 0pt{$\;\overline{\phantom{X}}$}\cM}}
\def\barcD{{\hfuzz=100cm\hbox to 0pt{$\;\overline{\phantom{X}}$}\cD}}
\def\ff#1#2{{\textstyle\frac{#1}{#2}}}

\def\spnh{\Spin(32)/\Z_2}

\def\HS#1{{\mathbb{F}}_{#1}}
\def\Ist#1{{\mathrm{I}\vphantom{\mathrm{II}}}^*_{#1}}
\def\mf#1{\mathfrak{#1}}

\begin{document}
\setcounter{page}0
\def\ppnumber{\vbox{\baselineskip14pt\hbox{DUKE-CGTP-98-03}
\hbox{NSF-ITP-98-027}\hbox{hep-th/9805206}}}
\def\ppdate{May, 1998} \date{}

\title{\LARGE Non-Simply-Connected Gauge Groups and\\
Rational Points on Elliptic Curves\\[10mm]}
\author{
Paul S. Aspinwall, David R. Morrison\\[10mm]
\normalsize Center for Geometry and Theoretical Physics, \\
\normalsize Box 90318, \\
\normalsize Duke University, \\
\normalsize Durham, NC 27708-0318\\[10mm]
}

{\hfuzz=10cm\maketitle}

\def\Large{\large}
\def\LARGE{\large\bf}


\begin{abstract}
We consider the F-theory description of non-simply-connected gauge groups
appearing in the $E_8\times E_8$ heterotic string. The analysis is closely
tied to the arithmetic of torsion points on an elliptic curve. The general
form of the corresponding elliptic fibration is given for all finite
subgroups of $E_8$ which are applicable in this context. We also study the
closely-related question of point-like instantons on a K3 surface whose
holonomy is a finite group.  As an example we consider the case of the
heterotic string on a K3 surface having the $E_8$ gauge symmetry broken to
$\SU(9)/\Z_3$ or $(E_6\times\SU(3))/\Z_3$ by point-like instantons with
$\Z_3$ holonomy.
\end{abstract}

\vfil\break


\section{Introduction}    \label{s:int}

The subject of nonperturbative gauge symmetry has always been central to the
study of string duality. The observation that the type IIA string acquires a
nonperturbative gauge symmetry when compactified on a singular K3
surface is an essential ingredient in formulating a duality between this
theory and the heterotic
string compactified on a 4-torus \cite{W:dyn}.

The geometry of elliptic fibrations is particularly useful in
analyzing the situations in which such gauge symmetries arise. This
manifests itself clearly when ``F-theory'' compactifications are
considered along the lines of \cite{MV:F,MV:F2}. The general idea is
that an F-theory model constructed from an elliptic K3 surface is dual to the
heterotic string compactified on a 2-torus, and that F-theory constructed
from an elliptic \CY\
threefold is dual to the heterotic string on a K3 surface. There is a
further duality in each case which arises upon further compactification on
a 2-torus, in which F-theory is replaced by the type IIA string.

One of the great strengths of the F-theory (or type IIA) picture is
that the geometry on the F-theory side may then be used to predict
nonperturbative gauge symmetries for the heterotic string. Thus we are
able to understand effects which are actually nonperturbative in both
pictures at the same time.

The focus of this paper concerns the geometry of the elliptic
fibrations of F-theory when the gauge symmetry in question is a
non-simply-connected gauge group. This has already been studied to a
small extent when the F-theory for the $\spnh$ heterotic string was
elucidated in \cite{AG:sp32}. Here we will be more systematic and describe the
general method for determining $\pi_1$ of the gauge group, at least in
the case of F-theory on a K3 surface.

Once we understand how to obtain a gauge group with nontrivial $\pi_1$
it turns out to be rather easy to understand the phenomenon of
``point-like instantons with discrete holonomy''.  This is specific to
the case of the heterotic string on a K3 surface, $S_H$.  Recall that
the heterotic string requires a bundle on $S_H$ in order to specify a
particular compactification. The simplest case to understand is when
this bundle corresponds to the point-like instantons of
\cite{W:small-i}. This was first applied in the context we will be
discussing in \cite{MV:F,MV:F2}.

A point-like instanton may be loosely considered to correspond to a situation
on the boundary of the moduli space of smooth bundles in which the
curvature of the Yang-Mills connection is zero everywhere except at
one point where it diverges. The one point is the location of the
instanton. Similarly one may have multiple point-like instantons for
which the curvature is concentrated at a set of points.

In the simplest case, a point-like instanton has completely trivial
holonomy (with respect to a Yang-Mills connection) around any loop in
$S_H$. (We exclude loops which pass
directly through the instanton itself.) The properties of such
point-like instantons have been studied at great length in
\cite{W:small-i,SW:6d,MV:F2,AG:sp32,BLPSSW:so32,me:sppt,DM:qiv,In:RG6,
AM:po,BI:6d1,BI:6d2}.

We will be able to extend these methods to the case where the
curvature of the instanton has been pushed into a point (or a set of
points) but some discrete remnant of the holonomy remains. This is
rather like the tangent bundle of a orbifold close to an orbifold
point. Such instantons have also been discussed
in some of the references above but mainly from the D-brane
perspective (see, for example, \cite{BLPSSW:so32}). Here we will
concern ourselves with the general features
of such instantons from the point of view of the elliptic fibration of
F-theory. We will restrict attention to the $E_8\times E_8$ heterotic
string although there is no particular reason why the same methods
should not also apply in the $\spnh$ case.

In order to analyze the elliptic fibrations that give rise to
non-simply-connected gauge groups we need to understand something
about ``torsion points on an elliptic curve''. We review this subject
in section \ref{s:ell}. We also give all the examples that are
required to study subgroups of $E_8$ in the context of F-theory.

In section \ref{s:T2} we prove the main result of the paper concerning
the relationship between $\pi_1$ of the gauge symmetry of a heterotic
string on a 2-torus and the \MW\ group of the elliptic fibration in
F-theory.

Finally in section \ref{s:K3} we discuss the point-like instantons
with discrete holonomy for the heterotic string on a K3 surface. This
is actually a very large subject and an analysis of all possibilities
would be a huge undertaking. We illustrate some of the possibilities
with a couple of examples.


\section{Some Arithmetic of Elliptic Curves}  \label{s:ell}

We shall require some basic knowledge of the arithmetic properties of
elliptic curves. While most of these results are standard (see, for
example, \cite{ST:rat,Silver:Ell}) we include the basic ideas here
as they have not been used commonly in string theory.

We begin with everything valued in the field $\C$ of complex numbers for
simplicity.
An elliptic curve $E$ is a smooth cubic curve in $\P^2$ with a marked
point, $O\in E$. One may also consider an elliptic curve to be simply
a 2-torus on which a point has been chosen. In the latter picture there is
clearly a group
action $\GU(1)^2$ of translations of the 2-torus. This group may be
identified with $E$ itself by mapping a point $P$ to a translation
that takes $O$ to $P$. Similarly if one has two points $P$ and $Q$
then one may define a group law by which $P+ Q$ is also a point
in $E$. Here $P+ Q$ is defined by translating $P$ by the element of
$\GU(1)^2$ associated to $Q$ or {\em vice versa}.

This group law is obvious enough if $E$ is pictured as a 2-torus but
how do we picture it for a cubic curve? The answer is as follows. A
generic line, $\P^1\subset\P^2$ will intersect the cubic at three
points. Thus given $P$ and $Q$ we may define a point $R$ by
the third point of collision of the line $PQ$ with $E$. Now we may
define $P+ Q$ as the third point of collision of the line $OR$
with $E$. It is easy to see that $O+ P=P$ as required but a little
more subtle to show that this group law is, in fact, associative.

In general we should take multiplicities into account when we do the
addition law. That is, tangent lines should generically count as
double intersections and tangent lines at a point of inflection should
be counted as a triple intersection.

An important property of this group law is that it acts within an
elliptic curve defined over any field $k$. That is, suppose we define
a cubic in $\P^2$ where the homogeneous coordinates of $\P^2$ and the
coefficients of the cubic lie in a field $k$. Then for any two points
$P$ and $Q$ in the elliptic curve, $P+Q$ is also in the elliptic
curve. For example we may take $k=\Q$, the rational numbers. In this
case the set of rational points in the elliptic curve form a group,
known as the \MW\
group. Clearly this group is abelian (since ``$+$'' is
commutative). The result of the celebrated Mordell--Weil theorem is
that it is finitely generated.

Any elliptic curve can be mapped to $\P^2$ (with homogeneous coordinates
$[x,y,z]$) in such a way that its equation is in ``Weierstrass form'':
\begin{equation}
  y^2z + a_1 xyz + a_3 yz^2= x^3 + a_2x^2z + a_4xz^2 + a_6z^3,
		\label{eq:W1}
\end{equation}
with the chosen point $O$ located at $[0,1,0]$. We will use
affine coordinates by setting $z=1$, and regard $O$ as the point ``at
infinity''. (Roughly speaking, $O$ is at $(x,y)=(0,\infty)$.)
It is easily seen to be a point of inflection in $E$. A
change of coordinates can always be used to put (\ref{eq:W1}) in its
reduced form\footnote{Unless $k$ is a field of characteristic 2 or 3!  But
we will have no reason to consider that case here.}
\begin{equation}
  y^2 = x^3 + ax + b.
		\label{eq:W2}
\end{equation}
We will often be interested in putting a particular point\footnote{$O$
and $(0,0)$ are {\em not\/} the same point!} at
$(x,y)=(0,0)$ and we will use the more general form (\ref{eq:W1}) with
$a_6=0$.

The problem we need to solve for use later in this paper is that of
writing down a general form of an elliptic curve on which a particular finite
subgroup $\Phi$ of the \MW\ group has been specified. For example, let us
consider the case
$\Phi\cong\Z_2$. Thus there is a point $P\in E$, such that
$P+P=O$. $P$ is said to be a ``2-torsion'' point of $E$. Running
through the above construction we see that the line passing through
$P$ twice must also pass through $O$. That is, the tangent line at $P$
passes through $O$ and so the slope, $dy/dx$, is infinite at $P$.

Putting $P$ at $(0,0)$ implies that $a_3=0$. We may impose $a_1=0$
without loss of generality and so
\begin{equation}
  y^2 = x(x^2 + a_2x + a_4),
\end{equation}
generically has $\Phi\cong\Z_2$. $\Phi$ may be larger for specific
values of $(a_2,a_4)$.

We now list the possibilities we will require in this paper. In
each case, we take a point $P$ corresponding to one of the generators of
our group (of maximal order $n$) and change coordinates to put $P$ at
$(x,y)=(0,0)$.  Imposing the condition that $P$ has order $n$ then
constrains the form of the Weierstrass equation.  We also list the
relationship between the coefficients of (\ref{eq:W2}) and the
discriminant $\Delta = 4a^3+27b^2$. See \cite{Ku:ell} for more details.
\begin{itemize}
  \item $\Phi\cong\Z_2$:
\begin{equation}
   y^2 = x(x^2 + a_2x + a_4),
\end{equation}
and
\begin{equation}
\begin{split}
  a &= a_4 - \ff13a_2^2\\
  b &= \ff1{27}a_2(2a_2^2 - 9a_4)\\
  \Delta &= a_4^2(4a_4-a_2^2)
\end{split}
\end{equation}
  \item $\Phi\cong \Z_3$:
\begin{equation}
  y^2 + a_1xy + a_3y = x^3,    \label{eq:W3}
\end{equation}
and
\begin{equation}
\begin{split}
  a &= \ff12a_1a_3 - \ff1{48}a_1^4\\
  b &= \ff14a_3^2 + \ff1{864}a_1^6 - \ff1{24}a_1^3a_3\\
  \Delta &= \ff1{16}a_3^3(27a_3-a_1^3).
\end{split}   \label{eq:W3D}
\end{equation}
  \item $\Phi\cong \Z_4$:
\begin{equation}
  y^2 + a_1xy + a_1a_2y = x^3+a_2x^3,
\end{equation}
and
\begin{equation}
\begin{split}
  a &= -\ff1{48}a_1^4 + \ff13 a_1^2a_2 - \ff13a_2^2\\
  b &= \ff1{864}(a_1^2-8a_2)(a_1^4-16a_1^2a_1-8a_2^2)\\
  \Delta &= -\ff1{16}a_1^2a_2^4(a_1^2-16a_2).
\end{split}
\end{equation}
  \item $\Phi\cong \Z_5$:
\begin{equation}
  y^2 + a_1xy +(a_1 - b_1)b_1^2\,y = x^3 + (a_1 - b_1)b_1\,x^2,
\end{equation}
and
\begin{equation}
\begin{split}
a &= \ff {1}{6}{a_{1}}b_{1}^{3}
 - {\displaystyle \ff {1}{48}} a_{1}^{4} + {\displaystyle
\ff {1}{3}}a_{1}^{2}b_{1}^{2} - {\displaystyle \ff {
1}{3}}b_{1}^{4} - {\displaystyle \ff {1}{6}}a_{1}^{3}b_{1}\\
b &=  \ff1{864}(a_1^2 -2a_1b_1 + 2b_1^2)(a_1^4 + 14a_1^3b_1 +
 26a_1^2b_1^2 -116a_1b_1^3+76b_1^4)\\
\Delta &={\displaystyle \ff {1}{16}}(a_{1}^{2} + 9\,a
_{1}{b_{1}} - 11b_{1}^{2})({a_{1}} - {b_{1}})^{5}b_{1}^{5}.
\end{split}
\end{equation}
  \item $\Phi\cong \Z_6$:
\begin{equation}
  y^2 + a_1xy + \ff1{32}(a_1-b_1)(3a_1+b_1)(a_1+b_1) =
	x^3 + \ff18(a_1-b_1)(a_1+b_1)x^2,
\end{equation}
and
\begin{equation}
\begin{split}
a &= \ff1{192}b_1(3a_1^3 - 3a_1^2b_1 -3 a_1b_1^2 - b_1^3)\\
b &= \ff1{110592}(3a_1^2 - 6a_1b_1 - b_1^2)(9a_1^4 -6a_1^2b_1^2 -
	24a_1b_1^3 - 11b_1^4)\\
\Delta &= \ff1{2^{24}}(a_1-5b_1)(3a_1+b_1)^2(a_1+b_1)^3(a_1-b_1)^6.
\end{split}
\end{equation}
  \item $\Phi\cong \Z_2\oplus\Z_2$:
\begin{equation}
  y^2 = x(x-b_2)(x-c_2),
\end{equation}
and
\begin{equation}
\begin{split}
a &= \ff13(b_2c_2 - b_2^2 - c_2^2)\\
b &= -\ff1{27}(b_2+c_2)(b_2-2c_2)(2b_2-c_2)\\
\Delta &= -b_2^2c_2^2(b_2-c_2)^2.
\end{split}
\end{equation}
  \item $\Phi\cong \Z_4\oplus\Z_2$:
\begin{equation}
  y^2+a_1xy -a_1(b_1^2-\ff1{16}a_1^2)y = x^3 - (b_1^2 - \ff1{16}a_1^2)x^2,
\end{equation}
and
\begin{equation}
\begin{split}
a &= -\ff1{768}a_1^4 - \ff7{24}a_1^2b_1^2-\ff13b_1^4\\
b &= \ff1{55296}(a_1^2+16b_1^2)(a_1^2-24a_1b_1+16b_1^2)
        (a_1^2+24a_1b_1+16b_1^2)\\
\Delta &= -\frac1{2^{16}}a_1^2b_1^2(a_1-4b_1)^4(a_1+4b_1)^4
\end{split}
\end{equation}
\end{itemize}

The subscripts for the coefficient parameters in the above have been chosen to
correspond to their degree under a rescaling symmetry $(x,y)\mapsto
(\lambda^2x,\lambda^3y)$. In each case there are two parameters
leading to an effective one parameter family when this rescaling is
taken into account.

The power of having formulated the above discussion over an arbitrary field
becomes apparent when we choose the field to be $k=\C(s)$, the field
of functions of the form $p(s)/q(s)$ where $p(s)$ and $q(s)$
are polynomials in some new variable $s$. This corresponds to a family
of elliptic curves $\pi:\cX\to D$, where the generic fibre is an
elliptic curve and $D$ is a complex line with coordinate $s$.\footnote{More
generally, we could consider the field of rational functions on an
arbitrary base manifold $D$ (if it has an algebraic structure), such as a
Riemann surface or a higher-dimensional algebraic variety.}  A
``rational'' point in the field $\C(s)$ is now a {\em section\/} of
the bundle $\pi:\cX\to D$.

Thus we may define the \MW\ group of a family of elliptic curves as
the group of sections. The group law is exactly that as explained
above.

Note that rational points on an elliptic curve over $\Q$ can
be trivially rewritten as rational points on an elliptic curve over
$\C(s)$. The converse is not true however. Mazur's theorem
\cite{Maz:M1,Maz:M2} asserts that only fifteen possibilities are allowed
for the torsion part of $\Phi$ for an elliptic curve over $\Q$. One
possibility which is not allowed is $\Phi\cong\Z_3\oplus\Z_3$. This
{\em is\/} allowed for an elliptic curve over $\C(s)$ however. In this
case one may put
\begin{itemize}
  \item $\Phi\cong \Z_3\oplus\Z_3$:
\begin{multline}
  y^2+a_1xy -\ff13(a_1+\omega b_1)(a_1+\omega^2b_1)b_1\,y\\
	= x^3  - (a_1-b_1)b_1\,x^2
	+ \ff13(a_1+\omega b_1)(a_1+\omega^2b_1)b_1^2\,x,
\end{multline}
and
\begin{equation}
\begin{split}
a &= -\ff1{48}a_1(a_1-2b_1)(a_1-2\omega b_1)(a_1-2\omega^2b_1)\\
b &= \ff1{864}(a_1^2 + 2a_1b_1 -2b_1^2)
	(a_1^2 + 2\omega a_1b_1 -2\omega^2b_1^2)
	(a_1^2 + 2\omega^2a_1b_1 -2\omega b_1^2)\\
\Delta &= \ff1{432}(a_1+b_1)^3(a_1+\omega b_1)^3(a_1+\omega^2b_1)^3b_1^3,
\end{split}
\end{equation}
\end{itemize}
where $\omega$ is a nontrivial cube cube of unity.
Here one of the $\Z_3$ generators may be put at $(0,0)$ but any
generator of the other $\Z_3$, e.g., $(\ff13(a_1-2b_1)b_1
-\ff{\omega}3(a_1+b_1)b_1,0)$, fails to lie in $\Q$.


\section{The Heterotic String on a Two-Torus} \label{s:T2}

We begin with a quick statement of the elements of F-theory we
require. For a more thorough description see
\cite{MV:F,MV:F2,me:lK3,FMW:F,AM:po} and in particular
\cite{me:hyp}. We will treat the duality between F-theory
modeled on a space $Y$ and the heterotic string compactified on a space
$Z$ as a limit of the duality between the type IIA string compactified on
$Y$ and the heterotic string compactified on $Z\times T^2$.

In this section we consider F-theory on a K3 surface, $S_F$. This is
dual to the $E_8\times E_8$ heterotic string on a 2-torus. Of
particular interest is the case where this 2-torus is very large. In
this case $S_F$ degenerates into a reducible variety consisting of two
rational elliptic surfaces, $R_1$ and $R_2$, intersecting along an
elliptic curve, $E_*$. The complex structure of $E_*$ is identified with
that of the heterotic 2-torus \cite{FMW:F}. A similar picture for the
$\spnh$ heterotic string was constructed in \cite{AM:po}. (See also
\cite{CD:F4}.)

The geometry of these rational elliptic surfaces is of central
importance to us. Roughly speaking, each of $R_1$ and $R_2$ is
identified with one of the $E_8$ gauge groups of the heterotic
string. In particular, holding $E_*$ fixed,  deforming the
complex structure on $R_i$ is dual to deforming the corresponding $E_8$
bundle on $T^2$.

Let the proper structure group of the $E_8$-bundle on $T^2$ be
$\cG\subset E_8$. The observed gauge group in the heterotic string
theory from this primordial $E_8$ gauge symmetry will then be the
centralizer of $\cG\subset E_8$.

Dual to this, on the F-theory side, the gauge symmetry is produced by
2-spheres shrinking down to zero area within $R_i$. Let us consider
this from the elliptic fibration $\pi:R_i\to f$, where $f$ is a $\P^1$
and the generic fibre is an elliptic curve. This fibration has at
least one section which we label $\sigma_0$.
The elements of $H_2(R_i,\Z)$ corresponding to $\sigma_0$ and the
elliptic fibre form a 2-dimensional unimodular even sublattice,
$U\subset H_2(R_i,\Z)$. The fact that $H_2(R_i,\Z)$ is itself even and
unimodular of rank 10 shows that $H_2(R_i,\Z)\cong\Gamma_8\oplus U$,
where $\Gamma_8$ is
isomorphic to the root lattice of $E_8$.\footnote{The intersection
form in $H_2(R_i,\Z)$ gives $\Gamma_8$ a negative signature. In our
discussion of Lie algebras we will implicitly assume this signature
was positive. Although this is sloppy it helps simplify the discussion.}
In order to analyze how
2-spheres can shrink down we need the following \cite{MirPer:ell}

\begin{theorem}
Let $M$ be the lattice of 2-cycles
within the fibres of the elliptic fibration $\pi:R_i\to f$ which do
{\em not\/} intersect $\sigma_0$. Then
\begin{equation}
  0 \to M \to \Gamma_8 \to \Phi \to 0,  \label{eq:ES1}
\end{equation}
where $\Phi$ is the \MW\ group of this elliptic fibration.
\end{theorem}

The F-theory rule for analyzing the gauge group is that all components
in all the fibres of $\pi:R_i\to f$ should be shrunk to zero size. The
subset of elements of $\Gamma_8$, of self-intersection $-2$, which are
shrunk to zero size in this process form the roots of the observed
gauge symmetry {\em algebra\/}. This subset is given precisely by the
generators of $M$. That is, $M$ is the root lattice of the gauge
algebra.

In the generic case, all of the fibres of $\pi:R_i\to f$ are fibres of
Kodaira type I$_1$ and thus $M$ is trivial. In this case $\Phi$ is of
rank 8 and generates all of $\Gamma_8$. No elements of $H_2(R_i,\Z)$
corresponding to roots of $\Gamma_8$ are
shrunk to zero
size when the fibres are shrunk and so the gauge algebra is trivial. The
structure group of the bundle really is the full $E_8$.

At the other extreme, if one acquires a type II$^*$ fibre then
$M\cong\Gamma_8$ and $\Phi$ becomes trivial. The gauge algebra is $\mf{e}_8$
and the bundle has trivial holonomy. These are point-like instanton
solutions \cite{MV:F}.

We would like to know the gauge {\em group\/}, $\cG$, as well as the gauge
algebra for the F-theory picture.
We will do this by generalizing a method used for analysis of the
$\spnh$ heterotic string in \cite{AG:sp32}.
We require more than just a
knowledge of the massless vector states which give the gauge
bosons. We also need to analyze the representations of massive states
with respect to $\cG$.

To do this consider the BPS solitons of the type IIA string. These
correspond to 2-branes wrapped over elements of $H_2(R_i,\Z)$. The
Ramond-Ramond charges of these states are given by this homology
class. This allows one to determine exactly the representation theory for
the BPS states when we have an enhanced gauge symmetry.

Given the exact sequence (\ref{eq:ES1}) we may use standard methods in
algebra (see, for example, chapter 3 of \cite{MacL:hom}) to yield
\begin{equation}
  0 \to \Hom(\Phi,\Z) \to \Hom(\Gamma_8,\Z) \mathrel{\mathop\to^j} \Hom(M,\Z)
	\to \Ext(\Phi,\Z)\to 0.   \label{eq:ES2}
\end{equation}
Now $\Phi$ may be decomposed into its free part $\Free(\Phi)$ and its
torsion part $\Tors(\Phi)$. We have $\Hom(\Phi,\Z)\cong\Free(\Phi)$
and $\Ext(\Phi,\Z)\cong\Tors(\Phi)$. Also
$\Hom(\Gamma_8,\Z)\cong\Gamma_8$ and $\Hom(M,\Z)$ is the {\em weight
lattice\/} of the algebra of $\cG$.

\begin{table}
\begin{center}
\begin{tabular}{|c|c|c|}
\hline
Finite group&Centralizer&Kodaira Fibres\\
\hline
$\{\text{id}\}$ & $E_8$ & II$^*$, 2I$_1$ {\it or}\/ II$^*$, II \\
$\Z_2$ & $\Spin(16)/\Z_2$ & $\Ist4$, 2I$_1$ \\
$\Z_2$ & $(E_7\times\SU(2))/\Z_2$ & III$^*$, I$_2$, I$_1$ {\it or}\/
            III$^*$, III\\
$\Z_3$ & $\SU(9)/\Z_3$ & I$_9$, 3I$_1$ \\
$\Z_3$ & $(E_6\times\SU(3))/\Z_3$ & IV$^*$, I$_3$, I$_1$ {\it or}\/ IV$^*$,
           IV\\
$\Z_4$ & $(\SU(8)\times\SU(2))/\Z_4$ & I$_8$, I$_2$, 2I$_1$\\
$\Z_4$ & $(\Spin(10)\times\Spin(6))\Z_4$ & $\Ist1$, I$_4$, I$_1$\\
$\Z_5$ & $(\SU(5)\times\SU(5))/\Z_5$ & 2I$_5$, 2I$_1$\\
$\Z_6$ & $(\SU(6)\times\SU(3)\times\SU(2))/\Z_6$ & I$_6$, I$_3$,
	I$_2$, I$_1$ \\
$\Z_2\oplus\Z_2$ & $(\Spin(12)\times\Spin(4))/(\Z_2\times\Z_2)$
	& $\Ist2$, 2I$_2$\\
$\Z_2\oplus\Z_2$ & $(\Spin(8)\times\Spin(8))/(\Z_2\times\Z_2)$
	& 2$\Ist0$\\
$\Z_4\oplus\Z_2$ & $(\SU(4)^2\times\SU(2)^2)/(\Z_4\times\Z_2)$
	& 2I$_4$, 2I$_2$\\
$\Z_3\oplus\Z_3$ & $\SU(3)^4/(\Z_3\times\Z_3)$
	& 4I$_3$\\
\hline
\end{tabular}
\end{center}
\caption{Finite abelian subgroups of $E_8$ with at most two generators,
their centralizers, and the
Kodaira fibres of the corresponding rational elliptic surface.}
\label{tab:1}
\end{table}

Of particular interest is the map $j$ in (\ref{eq:ES2}). This is
mapping the RR-charges of the BPS states into the weight lattice of
the gauge algebra. If the gauge group is not simply connected we
expect certain representations of the algebra to be missing. This is
measured by the cokernel of $j$ which is isomorphic to
$\Ext(\Phi,\Z)\cong\Tors(\Phi)$. In fact, a little group theory yields
the following
\begin{theorem}
  If the BPS solitons form a faithful representation of the gauge
group $\cG$, then
\begin{equation}
  \pi_1(\cG) \cong \Tors(\Phi).
\end{equation}
\end{theorem}

There is a further check that could be made of our claim that $\pi_1(\cG)$
is non-trivial --- we could analyze the behavior of the theory on
non-trivial spacetimes.  We will not attempt to do this in detail, but just
point out that when the theory is further compactified on $T^2$, we would
expect to see $\pi_1(\cG)$ as a symmetry of the theory.  And indeed, as a
subgroup of the \MW\ group it acts on the compactified IIA string.

Fortunately all possible forms of the rational elliptic surface as an
elliptic fibration have been listed by Persson \cite{Pers:RES} and
$\Phi$ has been determined in each case.  In particular, the list of
possible torsion groups is completely known.  It appears in the first
column of table~1.

Each of these torsion groups can be embedded in $E_8$, with one or two
inequivalent embeddings in each case.  (The equivalence is conjugacy within
$E_8$.)  In fact, this is the complete list of conjugacy classes of finite
abelian subgroups of
$E_8$ with at most two generators.\footnote{More than two generators would
not be appropriate on the heterotic side, since we only have two Wilson
lines available to generate the holonomy.}  If we calculate the centralizer
of each embedded torsion group, we
will determine the maximum gauge group $\cG$ for each possible $\pi_1$.
This is done in the second column of the table.  All of the corresponding
groups have rank 8.  And in fact, each of these
possibilities is realized in F-theory, due to the classification of rational
elliptic surfaces of this type given in \cite{MirPer:ell}.  We list the
Kodaira fibres of the corresponding rational surfaces\footnote{In a few
cases, two surfaces are possible --- the one with the I$_k$ fibres is
generic.}  in the third column
of table 1.

For example, consider one of the rational elliptic surfaces listed in
\cite{MirPer:ell,Pers:RES} which consists of a fibration with Kodaira
fibres of the class one of I$_9$ and three of I$_1$. According to the
usual rules of F-theory, the gauge algebra associated to the I$_9$
fibre will be $\su(9)$. Persson tells us that $\Phi\cong\Z_3$ in this
case however and so $\cG\cong\SU(9)/\Z_3$.

Note that $\SU(9)/\Z_3$ is a centralizer of $\Z_3\subset E_8$ and so
this corresponds to a heterotic string on a bundle with holonomy
$\Z_3$. That is, one of the Wilson loops of the heterotic 2-torus is
such that going around it three times is trivial.

Note that in each case $\pi_1(\cG)$ is actually
isomorphic to the discrete centralizer. This is a special property of
$E_8$, due to the $E_8$ lattice being unimodular.


\section{The Heterotic String on a K3 Surface} \label{s:K3}

Statements about duality become more interesting when we consider
F-theory on a \CY\ threefold, $X$, and the dual picture of a
$E_8\times E_8$ heterotic string on a K3 surface, $S_H$.
Since K3 is simply-connected we will need point-like instantons to
generate the finite holonomy of the preceding section.

To do the analysis we
may write $S_H$ as an elliptic fibration $\pi_H:S_H\to B$, where
$B\cong\P^1$ and apply the duality of the preceding section
``fibre-wise''.
We will assume $X$ may be written in the form of a K3 fibration $p:X\to B$ and
an elliptic fibration $\pi_F:X\to\Sigma$ with at least one section,
$\sigma_0$. See \cite{me:lK3} for a description of the conditions
required for this assumption.

As $S_H$ becomes large, $X$ undergoes a stable degeneration in which
each generic K3 fibre of the map $p$, becomes a union $R_1\cup_{E_*}
R_2$ of two rational elliptic surfaces intersecting along an elliptic
curve just as in the last section. $S_H$ may then be identified as the
elliptic fibration $\pi_H:S_H\to B$ with fibre isomorphic to $E_*$.

Again gauge symmetry is produced by 2-cycles shrinking down within a generic
$R_i$ when the elliptic fibre of the fibration $\pi_F:X\to\Sigma$ is
shrunk down to zero size. Globally this corresponds to 4-cycles
collapsing onto 2-cycles within $X$.

This is most easily represented by thinking of the discriminant locus,
$\Delta$, of the fibration $\pi_F:X\to\Sigma$. $\Delta$ corresponds to
a divisor within $\Sigma$. The degeneration of the fibre of Kodaira
type worse than I$_1$ or $II$ along a component of $\Delta$ will give
rise to an enhanced gauge symmetry. See \cite{MV:F,MV:F2,me:lK3} for
more details.

$\Sigma$ may also be written as a $\P^1$-fibration over $B$. Let $f$
denote a generic $\P^1$ fibre of this fibration. Thus $R_i$ is an
elliptic fibration of $f$ as in the previous section. We may restrict
the fibration $\pi_F:X\to B$ to each $f$ and look at the \MW\ group of
each rational elliptic surface as in the previous section. Again, any torsion
here will cause an obstruction to certain BPS states forming any
representation of the gauge algebra.

A new aspect arises over singularities of $\Delta$ such as when
irreducible components of $\Delta$
collide. The fibre over such a singularity (which need not be within
Kodaira's classification) may contain 2-cycles whose homology
class did not appear in nearby fibres. The result of wrapping 2-branes
over such cycles results in hypermultiplets. Thus, in order to find
which representations of the gauge group appear we should extend the
analysis of the previous section to such singularities within
$\Delta$. The analysis of the \MW\ group is much more difficult in this
case (due to possibilities such as ``birational sections'' which were not
an issue in lower dimension), but it seems likely that a result similar to
the conclusion of theorem~2 still holds, namely, that  $\pi_1(\cG)\cong
\Tors(\Phi)$ where $\Phi$ is the \MW\ group.  We will restrict our
attention to models for which this does hold.

The possibilities one may analyze are extremely numerous and there are
many interesting features about most of the possibilities. Here we
will just give an example to give the flavour of the
subject. The Weierstrass forms which would be necessary to compute any
example are listed in section \ref{s:ell}.

The method we use is standard to F-theory and comes from
\cite{MV:F2}. The notation we use follows
\cite{AG:sp32,AM:po,me:hyp}. In particular we take $\Sigma$ to be the
Hirzebruch surface $\HS n$ and $C_0$ to be a $\P^1$ section of
$\Sigma$ fibred over $B$ with self-intersection $-n$. In the heterotic
string picture this describes an $E_8$-bundle with $c_2=12-n$. The
methods used to analyze the geometry are very similar to those
appearing in the above references and for that reason we focus mainly
on the results rather than details of the method.

\subsection{$\Z_3$ Point-like Instantons}   \label{ss:Z3}

Suppose we have a point-like instanton on a K3 surface whose holonomy
is $\Z_3$. This only makes sense if the topology of the boundary of a
small neighbourhood around the instanton has $\pi_1\supset\Z_3$. One
way to arrange this is to have the instanton located on a $\C^2/\Z_3$
singularity of the K3 surface. In this case the boundary of a small
neighbourhood around the instanton is a lens space $S^3/\Z_3$.

We will just focus on one of the $E_8$'s of the heterotic string. We
expect $E_8$ to be broken to the centralizer of $\Z_3\subset
E_8$. There are two possibilities for embedding $\Z_3$ in $E_8$ and thus
two possible centralizers. Each has quite different geometry and thus
physics.

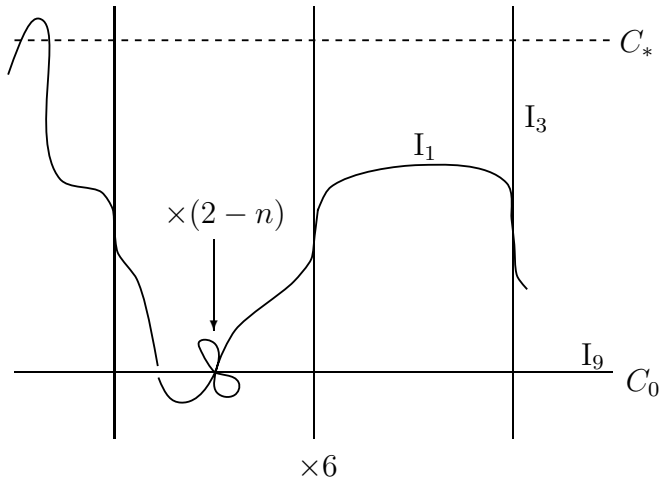
\begin{figure}
\begin{center}
\setlength{\unitlength}{0.00058300in}%
\begin{picture}(5586,4287)(1435,-4336)
\thinlines
\put(1501,-3361){\line( 1, 0){5400}}
\put(2401,-61){\line( 0,-1){3900}}
\put(4201,-61){\line( 0,-1){3900}}
\put(6001,-61){\line( 0,-1){3900}}
\put(1501,-361){
}
\put(3301,-2161){\vector( 0,-1){825}}
\put(1443,-673){
}
\put(2801,-3410){
}
\put(4238,-1904){
}
\put(6616,-3292){\makebox(0,0)[lb]{\smash{I$_9$}}}
\put(6961,-472){\makebox(0,0)[lb]{\smash{$C_*$}}}
\put(7021,-3532){\makebox(0,0)[lb]{\smash{$C_0$}}}
\put(5101,-1402){\makebox(0,0)[lb]{\smash{I$_1$}}}
\put(6106,-1147){\makebox(0,0)[lb]{\smash{I$_3$}}}
\put(4056,-4297){\makebox(0,0)[lb]{\smash{$\times6$}}}
\put(2851,-2011){\makebox(0,0)[lb]{\smash{$\times(2-n)$}}}
\end{picture}
\end{center}
  \caption{A perturbative gauge group of $\SU(9)/\Z_3$.}
\label{fig:I9}
\end{figure}

One possibility is $\SU(9)/\Z_3$. We know that because of the $\Z_3$,
the Weierstrass form must be given by (\ref{eq:W3}). We may achieve
the curve of I$_9$ fibres along $C_0$ required to generate the
$\su(9)$ symmetry by making $a_3$ vanish to order 3 along this
curve. The resulting discriminant is then forced into the form shown
in figure \ref{fig:I9}.

The first thing to note is that six lines of I$_3$ fibres appear in
the $f$ direction. Each such line crosses $C_*$. Recall that $C_*$ is
the section of $S_H$ as an elliptic fibration. This implies that {\em $S_H$
generically has six quotient singularities of the form $\C^2/\Z_3$.}

Thus the F-theory picture is in nice agreement with the heterotic
picture. In order to get an $\SU(9)/\Z_3$ gauge symmetry we must have
fractionally charged instantons and thus $S_H$ must have quotient
singularities. The key to this happening in the F-theory picture is
that the form of the discriminant \ref{eq:W3D}) has a factor which is
a perfect cube. Note that all of the forms of the discriminant listed
in section \ref{s:ell} have strong factorization properties and so
this is a feature which is common to all examples.  Actually F-theory
also managed to correctly count the number of orbifold points in $S_H$
as we now argue.

The bundle, $V\to S_H$ for the heterotic string has holonomy $\Z_3$. This
implies that there is a map $\phi:S_C\to S_H$ which is three-to-one
such that $\phi^*V$ has trivial holonomy. $\phi$ will have fixed
points at the location of the orbifold points. Let the number of
orbifold points be $m$. Given that $S_C$ is a K3 surface\footnote{One
might also think that $S_C$ could be a 4-torus. However, in that case
$S_H$ would not be the Weierstrass model of an elliptic surface with a
section.}
and that $S_H$ is a K3
surface, one may do a simple Euler characteristic calculation to
determine $m$:
\begin{equation}
  \frac{24-m}{3} + 3m = 24.
\end{equation}
Thus $m=6$ in agreement with F-theory.

As well as producing a $\C^2/\Z_3$ orbifold point in $S_H$, each
vertical line of I$_3$ fibres also produces a nonperturbative gauge
symmetry of $\su(3)$. Thus, a $\C^2/\Z_3$ orbifold point with this
fractional point-like instanton produces a nonperturbative gauge
enhancement of $\su(3)$. The collisions of the vertical lines of I$_3$
fibres with the line of I$_9$ fibres along $C_0$ each produce
hypermultiplets in the $(\mathbf{3},\mathbf{9})$ representation of
$\su(3)\oplus\su(9)$. 
The result is that the gauge group is
\begin{equation}
  \cG \cong\frac{\SU(9)\times\SU(3)^{6}}{\Z_3}.
\end{equation}

The remaining I$_1$ part of the discriminant collides with degree 3 with the
lines of I$_3$ fibres and I$_9$ fibres. The collisions with I$_3$
lines produce nothing interesting but the collisions with $C_0$ at a
total of $2-n$ points each produce a massless tensor.
Using arguments along the lines of \cite{me:hyp} one may analyze the
moduli space of the object associated with each tensor. One finds that
it corresponds to a copy of $S_H$. That is, it is a free object
allowed to be anywhere in $S_H$.
In fact, these
massless tensors are essentially identical to the usual point-like
$E_8$ instantons with trivial holonomy of \cite{SW:6d,MV:F}. Such
point-like instantons are known to have $c_2=1$.
Since the total $c_2$ of this bundle should be $12-n$, the fractional
point-like instantons appear to each contribute
$((12-n)-(2-n))/6=\ff53$ towards $c_2$.

\begin{figure}
\begin{center}
\setlength{\unitlength}{0.00058300in}%
\begin{picture}(5532,3679)(1489,-3704)
\thinlines
\put(1501,-3361){\line( 1, 0){5400}}
\put(1501,-361){
}
\put(6301,-1261){\vector( 0, 1){750}}
\put(4576,-1711){\vector(-1, 0){300}}
\put(1951,-2461){\vector( 0,-1){750}}
\put(1576,-2311){
}
\put(1771,-3539){
}
\put(3765,-3284){
}
\put(6961,-472){\makebox(0,0)[lb]{\smash{$C_*$}}}
\put(7021,-3532){\makebox(0,0)[lb]{\smash{$C_0$}}}
\put(6466,-3292){\makebox(0,0)[lb]{\smash{IV$^*$}}}
\put(6751,-1036){\makebox(0,0)[lb]{\smash{I$_3$}}}
\put(6226,-1486){\makebox(0,0)[lb]{\smash{$\times6$}}}
\put(4726,-1861){\makebox(0,0)[lb]{\smash{$\times2$}}}
\put(1876,-2311){\makebox(0,0)[lb]{\smash{$\times(6-n$)}}}
\put(3976,-3061){\makebox(0,0)[lb]{\smash{I$_1$}}}
\end{picture}
\end{center}
  \caption{A perturbative gauge group of $(E_6\times\SU(3))/\Z_3$.}
\label{fig:IV*}
\end{figure}
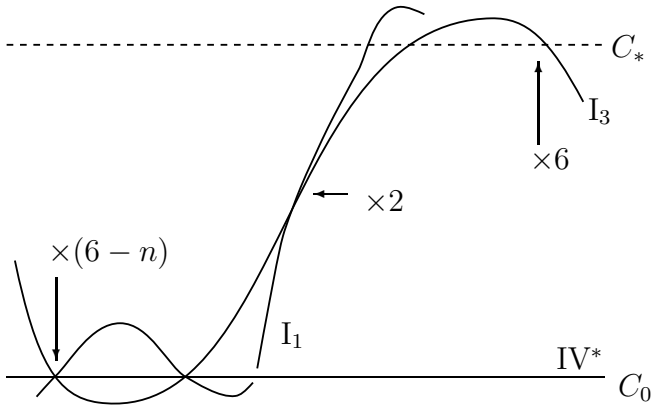

As usual, one may check the anomalies of this six-dimensional theory
(see, for example, \cite{Sch:a6}). When doing this it is important to
remember that by going to the stable degeneration we are only looking
at ``half'' of the $E_8\times E_8$ heterotic string. The anomalies will
only cancel if the {\em full\/} massless particle spectrum is determined.

Let us now consider another example.
$\Z_3$ may be embedded in $E_8$ in an inequivalent way
such that it centralizes $(E_6\times\SU(3))/\Z_3$. We may put a line
of type IV$^*$ fibres along $C_0$ to achieve this. The result is shown
in figure \ref{fig:IV*}.

The curve of I$_3$ fibres crosses $C_*$ six times. Thus the
heterotic K3 surface, $S_H$, is forced to have six $\C^2/\Z_3$
orbifold points again. Beyond this however, this situation seems
altogether milder than the $\SU(9)/\Z_3$ case above. There are no new
nonperturbative gauge groups. There are $(6-n)$ blow-ups in the base
implying $6-n$ old-fashioned point-like $E_8$ instantons. Therefore each
fractional instanton contributes $c_2=1$ to get
a total $c_2=12-n$.

Note that we can obtain more possibilities from these two $\Z_3$
examples. One may allow some of the point-like instantons with trivial
holonomy to coalesce with the orbifold points containing the
fractional point-like instantons. This yields many more
nonperturbative gauge groups, tensors and hypermultiplets. The
situation is very similar to that described in \cite{AM:po} so we will
not pursue it further here.


\section*{Note added}

When a specific torsion subgroup $\Phi$ is imposed on all elliptic curves in a
family, as happened in the situations studied in this paper, the monodromy
group of the family of elliptic curves is reduced from $SL(2,\Z)$ to a
subgroup $\Gamma$ of finite index.  In the case $\Phi=\Z_n$, the
corresponding monodromy
group is known as $\Gamma_1(n)$, and in the case $\Phi=\Z_n\times \Z_n$, the
corresponding monodromy group is known as $\Gamma(n)$.  There is a closely
related monodromy group $\Gamma_0(n)$ corresponding to a cyclic subgroup in
which a generator has not been chosen.

As this paper was being put into final form, two interesting papers
appeared\cite{BPS:Fquan,BKMT:IIBv} which study F-theory models with reduced
monodromy
$\Gamma_0(n)$ and $\Gamma(n)$.  The analysis given here concerns
conventional F-theory compactifications using such families of elliptic
curves, whereas \cite{BPS:Fquan,BKMT:IIBv} consider unconventional
compactifications in which an additional field has been turned on.

\section*{Acknowledgements}

It is a pleasure to thank R.~Bryant, R.~Hain, R.~Plesser,
D.~Reed, and N.~Seiberg for useful conversations, and to acknowledge
the hospitality and support of the Institute for Theoretical Physics,
U.C. Santa Barbara, where part of this work was done.
The work of D.R.M.\ is supported in part by
by NSF grant DMS-9401447.


\begin{thebibliography}{10}

\bibitem{W:dyn}
E.~Witten,
\newblock {\em String Theory Dynamics in Various Dimensions},
\newblock Nucl. Phys. {\bf B443} (1995) 85--126, hep-th/9503124.

\bibitem{MV:F}
D.~R. Morrison and C.~Vafa,
\newblock {\em Compactifications of F-Theory on Calabi--Yau Threefolds --- I},
\newblock Nucl. Phys. {\bf B473} (1996) 74--92, hep-th/9602114.

\bibitem{MV:F2}
D.~R. Morrison and C.~Vafa,
\newblock {\em Compactifications of F-Theory on Calabi--Yau Threefolds --- II},
\newblock Nucl. Phys. {\bf B476} (1996) 437--469, hep-th/9603161.

\bibitem{AG:sp32}
P.~S. Aspinwall and M.~Gross,
\newblock {\em The SO(32) Heterotic String on a K3 Surface},
\newblock Phys. Lett. {\bf B387} (1996) 735--742, hep-th/9605131.

\bibitem{W:small-i}
E.~Witten,
\newblock {\em Small Instantons in String Theory},
\newblock Nucl. Phys. {\bf B460} (1996) 541--559, hep-th/9511030.

\bibitem{SW:6d}
N.~Seiberg and E.~Witten,
\newblock {\em Comments on String Dynamics in Six Dimensions},
\newblock Nucl. Phys. {\bf B471} (1996) 121--134, hep-th/9603003.

\bibitem{BLPSSW:so32}
M.~Berkooz et~al.,
\newblock {\em Anomalies, Dualities, and Topology of $D=6$ $N=1$ Superstring
  Vacua},
\newblock Nucl. Phys. {\bf B475} (1996) 115--148, hep-th/9605184.

\bibitem{me:sppt}
P.~S. Aspinwall,
\newblock {\em Point-like Instantons and the $\Spin(32)/\Z_2$ Heterotic
  String},
\newblock Nucl. Phys. {\bf B496} (1997) 149--176, hep-th/9612108.

\bibitem{DM:qiv}
M.~R. Douglas and G.~Moore,
\newblock {\em D-branes, Quivers, and ALE Instantons},
\newblock hep-th/9603167.

\bibitem{In:RG6}
K.~Intriligator,
\newblock {\em RG Fixed Points in Six Dimensions via Branes at Orbifold
  Singularities},
\newblock Nucl. Phys. {\bf B496} (1997) 177--190, hep-th/9702038.

\bibitem{AM:po}
P.~S. Aspinwall and D.~R. Morrison,
\newblock {\em Point-like Instantons on K3 Orbifolds},
\newblock Nucl. Phys. {\bf B503} (1997) 533--564, hep-th/9705104.

\bibitem{BI:6d1}
J.~D. Blum and K.~Intriligator,
\newblock {\em Consistency Conditions for Branes at Orbifold Singularities},
\newblock Nucl. Phys. {\bf B506} (1997) 223--235, hep-th/9705030.

\bibitem{BI:6d2}
J.~D. Blum and K.~Intriligator,
\newblock {\em New Phases of String Theory and 6d RG Fixed Points via Branes at
  Orbifold Singularities},
\newblock Nucl. Phys. {\bf B506} (1997) 199--222, hep-th/9705044.

\bibitem{ST:rat}
J.~H. Silverman and J.~Tate,
\newblock {\em Rational Points on Elliptic Curves},
\newblock Undergraduate Texts in Mathematics, Springer-Verlag, 1992.

\bibitem{Silver:Ell}
J.~H. Silverman,
\newblock {\em The Arithmetic of Elliptic Curves}, Graduate Texts in
  Mathematics~{\bf 106},
\newblock Springer-Verlag, 1991.

\bibitem{Ku:ell}
D.~Kubert,
\newblock {\em Universal Bounds on the Torsion of Elliptic Curves},
\newblock Proc. London Math. Soc. {\bf 33} (1976) 193--237.

\bibitem{Maz:M1}
B.~Mazur,
\newblock {\em Modular Curves and Eisenstein Ideal},
\newblock IHES Publ. Math. {\bf 47} (1977) 33--186.

\bibitem{Maz:M2}
B.~Mazur,
\newblock {\em Rational Isogenies of Prime Degree},
\newblock Invent. Math. {\bf 44} (1978) 129--162.

\bibitem{me:lK3}
P.~S. Aspinwall,
\newblock {\em K3 Surfaces and String Duality},
\newblock in C.~Esthimiou and B.~Greene, editors, ``Fields, Strings and
  Duality, TASI 1996'', pages 421--540, World Scientific, 1997,
\newblock hep-th/9611137.

\bibitem{FMW:F}
R.~Friedman, J.~Morgan, and E.~Witten,
\newblock {\em Vector Bundles and F Theory},
\newblock Commun. Math. Phys. {\bf 187} (1997) 679--743, hep-th/9701162.

\bibitem{me:hyp}
P.~S. Aspinwall,
\newblock {\em Aspects of the Hypermultiplet Moduli Space in String Duality},
\newblock J. High Energy Phys. {\bf 04} (1998) 019, hep-th/9802194.

\bibitem{CD:F4}
G.~Curio and R.~Y. Donagi,
\newblock {\em Moduli in $N=1$ Heterotic/F-Theory Duality},
\newblock hep-th/9801057.

\bibitem{MirPer:ell}
R.~Miranda and U.~Persson,
\newblock {\em On Extremal Rational Elliptic Surfaces},
\newblock Math. Z. {\bf 193} (1986) 537--558.

\bibitem{MacL:hom}
S.~MacLane,
\newblock {\em Homology},
\newblock Springer, 1975.

\bibitem{Pers:RES}
U.~Persson,
\newblock {\em Configurations of Kodaira Fibres on Rational Elliptic Surfaces},
\newblock Math. Z. {\bf 205} (1990) 1--47.

\bibitem{Sch:a6}
J.~H. Schwarz,
\newblock {\em Anomaly-Free Supersymmetric Models in Six Dimensions},
\newblock Phys. Lett. {\bf B371} (1996) 223--230, hep-th/9512053.

\bibitem{BPS:Fquan}
M.~Bershadsky, T.~Pantev, and V.~Sadov,
\newblock {\em F-Theory with Quantized Fluxes},
\newblock hep-th/9805056.

\bibitem{BKMT:IIBv}
P.~Berglund, A.~Klemm, P.~Mayr, and S.~Theisen,
\newblock {\em On Type IIB Vacua With Varying Coupling Constant},
\newblock hep-th/9805189.

\end{thebibliography}

\end{document}